\documentclass[lettersize,journal]{IEEEtran}
\usepackage{amsmath,amsfonts}
\usepackage{algorithmic}
\usepackage{algorithm}
\usepackage{array}
\usepackage[caption=false,font=normalsize,labelfont=sf,textfont=sf]{subfig}
\usepackage{textcomp}
\usepackage{stfloats}
\usepackage{url}
\usepackage{verbatim}
\usepackage{graphicx}
\usepackage{cite}
\usepackage{amsmath,amssymb,amsfonts}
\usepackage{comment}
\usepackage{hhline}
\usepackage{float}
\usepackage{siunitx}

\begin{document}

\title{Double-Corrugated Degenerate Band-Edge Oscillator Driven by an Electron Beam}

\author{Miguel Saavedra-Melo, Robert Marosi, Nelson Castro, Eva Rajo-Iglesias, Filippo Capolino
\thanks{This material is based upon work partly supported by the Air Force Office of Scientific Research Multidisciplinary Research Program of the University Research Initiative (MURI) under grant number FA9550-20-1-0409 administered through the University of New Mexico. The work was partially funded by the Spanish Government under the grant PID2022-141055NB-C22 from MCIN/AEI/10.13039/501100011033 and in part by RED2022-134657-T.}
\thanks{Miguel Saavedra-Melo, Robert Marosi, and Filippo Capolino are with the Department of Electrical Engineering and Computer Science, University of California, Irvine, CA 92697 USA (e-mail: f.capolino@uci.edu).}
\thanks{Nelson Castro, Eva Rajo Iglesias, and Filippo Capolino and are with the Department of Signal Theory and Communications, University Carlos III of Madrid, Leganés, 28911, Spain (email: ncastro@pa.uc3m.es).}}

\maketitle

\begin{abstract}
The properties of a degenerate band-edge (DBE) oscillator based on the synchronous interaction between a linear electron beam synchronized with four degenerate electromagnetic modes are presented. The DBE in the cold slow-wave structure is implemented in a rectangular waveguide with a staggered double corrugation in it. At the DBE, four modes coalesce into a single degenerate mode, forming a very flat dispersion diagram with vanishing or very small group velocity. The DBE oscillator frequency is robust to variations in electron beam current, electrons' velocity, and cavity length. The DBE oscillator concept provided here is expected to be extendable to several other geometries as long as the slow wave structure supports a DBE.
\end{abstract}

\section{Introduction}

Vacuum electronic devices, such as traveling-wave tubes (TWTs), magnetrons, and backward-wave oscillators (BWOs)
are among the most widely used technologies for generating high-power microwaves, an increasingly important capability for applications requiring efficient and reliable radiation sources. These devices have become prominent thanks to their robust design and technological maturity \cite{PaoloniMillimeter2021,armstrong2023frontiers,parker2002vacuum,Booske2011Vacuum,anilkumar2022historical,gilmour2011klystrons,carter2018microwave}. The core operating principle of TWTs and BWOs lies in the distributed interaction and velocity synchronization between an electromagnetic mode supported by a slow-wave structure (SWS) and a linear electron beam \cite{pierce-twt50}. The strength of this beam-wave interaction (characterized primarily by the interaction impedance) is a key factor that fundamentally characterizes the achievable performance of modern vacuum electron devices, influencing parameters such as power conversion efficiency, operating bandwidth, and device compactness \cite{gewartowski65CH10}.

A well-established approach to improve this interaction is dispersion engineering, where the modal characteristics, such as phase velocity, field distribution, and interaction impedance, of the SWS are carefully tailored to enhance synchronization and interaction with an electron beam \cite{Chipengo2021Dispersion}. However, advancing to millimeter-wave and terahertz frequencies introduces new challenges. As the operating frequency increases, device dimensions shrink, and the limitations of conventional fabrication techniques become more severe \cite{Gamzina_Nano_CNC_2016}. Furthermore, power handling and the suppression of unwanted modes are increasingly difficult to manage at these frequencies. Therefore, innovative slow-wave designs with engineered dispersion profiles are essential for enabling coherent, high-efficiency radiation in the millimeter-wave and sub-terahertz regime.


Corrugated rectangular waveguide SWSs have been investigated for use in sub-THz vacuum electronic devices. Such SWSs consist of gratings inside rectangular waveguides \cite{McVey1994Analysis,Joe1994Wave,Zaginaylov2000Full-wave}. More recently, novel double-corrugated, offset double-corrugated, and photonic crystal-assisted corrugated waveguide SWSs have been developed for sub-THz BWOs, offering higher interaction impedance and simplifying fabrication due to their compatibility with modern microfabrication processes \cite{Mineo2010Corrugated,MineoDouble2010,LetiziaPhotonic2015,PaoloniTHzBackward2016,Basu2021}.
 
In this paper, we use the concept of degenerate band edge (DBE) to establish a degenerate coalescence of four eigenmodes that are supported by an SWS prior to the introduction of an electron beam \cite{figotin_gigantic_2005, figotin_frozen_2006, Figotin07SlowWaveRes, figotin_slow_2011}. Indeed, recent advances in the theory and implementation of DBE structures have laid the foundation for novel high-efficiency oscillators and electron-beam devices where the ``supersynchronization'' with the electron beam is achieved using four degenerate modes \cite{othman2016giantAmplif,othman2016theory,Othman16LowStarting,Abdelshafy2018Electron}. 

The occurrence of the DBE has been demonstrated experimentally at microwave frequencies for a handful of waveguide geometries. 
In early work by Chabanov \cite{chabanov2008strongly}, the resonant transmission properties of split band-edge (SBE) in a circular waveguide and in layered anisotropic mediums \cite{YargaDegenerate2008} were explored; the SBE is highly related to the DBE.
Later experimental demonstrations of the DBE have been explored in fully metallic circular waveguides with periodic inclusions \cite{Othman17ExpDem}, in coupled microstrip lines \cite{abdelshafy2019TAPexceptionalExper}, using metamaterial transmission lines implemented in microstrip \cite{Mealy2020GeneralConditions}, and in substrate-integrated waveguides (SIWs) \cite{Zheng22MTTSynthMeasDBE}. While most studies have been focusing on the properties of the DBE in linear systems, only a handful of studies have explored the role of the DBE in saturable systems, and that is what is done in the present work. 

In vacuum electronics, the concept of supersynchronism (coherent interaction between an electron beam and four degenerate modes to reduce starting current) was introduced in \cite{othman2016giantAmplif,  Othman16LowStarting, othman2016theory}, and further advanced in \cite{Chipengo2017Backward, Zuboraj2017Propagation,Abdelshafy2018Electron}.
The work in \cite{MacLachlan2022Efficient} and \cite{Maclachlan2024sub} expands the findings in \cite{Abdelshafy2018Electron}, and further demonstrates the DBE oscillator concept using a circular overmoded waveguide with walls covered by a quasi 2D periodic surface lattice achieving multi-megawatt peak power with efficiencies of 24\% and over 50\%, respectively, and highlights their scalability and relevance to THz applications and fusion energy.

In this work, we investigate a novel periodic SWS that exhibits a DBE in its cold dispersion relation, is relatively easy to fabricate using micromachining techniques, and operates at millimeter-wave frequencies. When the SWS is driven by an electron beam to make it a DBE oscillator, the device can efficiently produce significant output powers at millimeter-wave frequencies. The DBE occurs at a specific point in the dispersion diagram where four modes coalesce, resulting in zero group velocity and a significant enhancement of beam–wave interaction. We demonstrate the stability of a DBE-based oscillator against variations in beam current and voltage, and we assess its performance for different device lengths, highlighting the robustness and scalability of this oscillation principle. In this paper, we perform for the first time a comprehensive analysis of an oscillator based on the DBE interacting with a linear electron beam, accounting for the nonlinear interaction and saturation effect modeled with a particle-in-cell (PIC) simulator.

\section{Degenerate Band edge and Four-Mode Supersynchronization}

A DBE in SWSs occurs when four electromagnetic (EM) eigenmodes not only share identical eigenvalues (i.e., wavenumbers) but also collapse into a single eigenvector (i.e., polarization state) \cite{figotin_gigantic_2005, abdelshafy2019TAPexceptionalExper}. This phenomenon stands in contrast to conventional modal degeneracies, such as in circular waveguides, where two TE$_{11}$ modes have the same wavenumbers but orthogonal polarizations. In a DBE, both the eigenvalues and the corresponding eigenvectors merge, representing a 4th-order singularity in the eigenmode dispersion diagram.

\begin{figure}[htpb]
\centering
\includegraphics[width=0.9\linewidth]{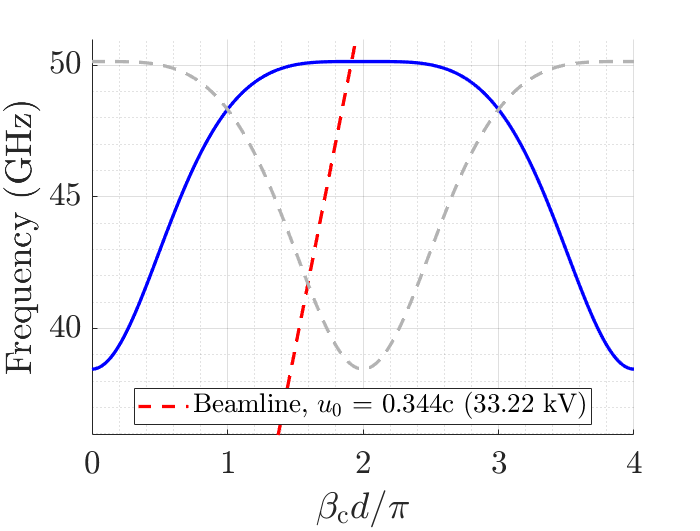}
\caption{The blue curve shows the typical very flat modal dispersion diagram of the modes in the cold SWS due to the DBE at the $\beta_{\mathrm{c}}d=2\pi$ point and $f_\mathrm{d} = 50.143$ GHz. The beamline, in dashed red, corresponds to a beam with electrons' velocity $u_0 = 0.344c$.}
\label{fig:disp_diagr_DBEO}
\end{figure}

In this work, we study a novel regime of operation for microwave oscillators, called ``four-mode supersynchronization". This regime is enabled by synchronizing the velocity of an electron beam with four EM modes that become degenerate under DBE conditions. The resulting synchronization drastically reduces the required beam current to initiate oscillation. 

The dispersion relation ($\beta_c$ vs. $f$) of the modes in the cold SWS, which will be described in the next section, is depicted in Fig. \ref{fig:disp_diagr_DBEO}. Here, $f_\mathrm{d}$ and $\beta_\mathrm{d}$ are the band-edge frequency and the wavenumber at which the DBE occurs in the cold SWS's dispersion diagram. The DBE condition requires the existence of at least four interacting modes. In Fig.~\ref{fig:disp_diagr_DBEO} only the propagating branches (in blue for the spatial harmonic of interest) for $f<f_\textrm{d}$ are displayed, while the evanescent branches are shown in the Appendix. In this paper, the SWS is periodic with period $d$, and $\beta_\mathrm{d} = 2\pi/d$. 
As $f$ approaches $f_\mathrm{d}$, all four modes coalesce,
and the dispersion relation near this point follows a characteristic quartic dependence:

\begin{equation}
    (f_\mathrm{d} - f) = \frac{h_\mathrm{d}}{2\pi} (\beta_\mathrm{c} - \beta_\mathrm{d})^4,
    \label{eq:dbe_relation}
\end{equation}

\noindent where $h_\textrm{d}$ is the parameter that describes the flatness of the dispersion. At the DBE, $d^n\omega/d\beta_\textrm{c}^n=0$ for $n=1,2,3$, and $d^4\omega/d \beta_\textrm{c}^4=- 24 h_\textrm{d}$. The presence of four coalescing modes that form a DBE, including the evanescent ones, is observed by considering the four quartic roots $(\beta_\textrm{c}-\beta_\textrm{d}) = \left[2\pi (f_\textrm{d}-f)/h_d\right]^{1/4}$ as clearly shown in \cite{abdelshafy2019TAPexceptionalExper, Mealy2020GeneralConditions}. When $f>f_\textrm{d}$, all four modes are evanescent for the case represented in Fig.~\ref{fig:disp_diagr_DBEO}. This concept is further demonstrated in the Appendix.

This unique dispersion profile leads to a point where the group velocity (and its derivatives) approaches zero, which results in significant field enhancement \cite{figotin_gigantic_2005}. These conditions are ideal for energy exchange with an electron beam, and stand in contrast to regimes associated with the regular band edge (RBE) discussed in \cite{HungAbsolute2015,AntoulinakisAbsolute2018}, where the Briggs–Bers criterion is applied to assess absolute instability, and in \cite{LiuCharacteristics2016}, where the formation of $3\pi$ stopbands in a V-band folded waveguide SWS is analyzed, demonstrating that geometric loadings play a key role in both stopband formation and the onset of instability. In \cite{du20203pi,shu2024study,zhao2016observation}, strong oscillations due to the beam-wave interaction at the band edge of stopbands in rectangular helix, double-staggered waveguide, and serpentine waveguide SWSs were also studied. The condition for four-mode supersynchronization is met when the average electron velocity $u_0$ of the beam satisfies

\begin{equation}
    u_0 \approx \frac{2 \pi f_\mathrm{d}}{\beta_\mathrm{d}}=f_\mathrm{d} d.
\end{equation}
As it will be shown later on in this paper, the DBE oscillator has a stable regime of oscillation also when the beam line intersects the dispersion diagram in the {\em neighborhood} of $\beta_\textrm{d}$ because of the large flatness of the dispersion due to the vanishing of the derivatives of the $\omega-k$ dispersion relation.

When an SWS with DBE has a finite length, the resulting cavity usually has a very high quality factor $Q$. Such a cold ``DBE cavity" has various resonances near the DBE frequency $f_\mathrm{d}$. Usually, the closest one to the DBE is called the ``DBE resonance", which becomes asymptotically closer to the DBE frequency when the number of unit cells $N$ of the cavity increases, 

\begin{equation}
    f_\textrm{r,d} \sim f_\textrm{d} - \nu_\textrm{d}/N^4,
    \label{eq:DBEresFreqAsymp}
\end{equation}

\noindent where $\nu_\textrm{d}= h_\textrm{d}(\pi-\varphi)^4/(2 \pi d^4)$, and $\sim$ denotes asymptotic equality for large $N$, as discussed in \cite{ burr2013OpExdegenerate, Othman17ExpDem, Mealy2020GeneralConditions}. Here, the angle $\varphi$ represents the reflection phase of the propagating waves at the edges of the cavity. Its precise value may be affected by the kind of termination, and it does not affect the key message of how fast $f_\textrm{r,d}$ asymptotically converges to $ f_\textrm{d}$ (note that the asymptotic formula reported in \cite{Othman16LowStarting, othman2016giantAmplif,Othman17ExpDem,abdelshafy2019TAPexceptionalExper} should have included the term $\varphi$ as was done in \cite{burr2013OpExdegenerate}). The resonance closest to the DBE is, in general, the sharpest one \cite{figotin_gigantic_2005}, and it is the one that couples the most to any small gain in a cavity; hence, it is the one that is responsible for self-sustained oscillations in this paper. 
Cavities based on DBE have a large quality factor, and when internal losses are neglected, the external quality factor grows asymptotically as $Q\propto N^5$ for large $N$, as demonstrated in \cite{Othman17ExpDem, Mealy2020GeneralConditions,Zheng22MTTSynthMeasDBE}. This property is robust with respect to variations in the load termination at the edge of the cavity because the DBE is generally mismatched to the ``outside world". Internal losses would slow the large $N^5$ growth of $Q$ at some large $N$ \cite{Mealy2020GeneralConditions}. Because of all these properties, the high gain arising from the four-wave synchronization is responsible for the self-sustained oscillations at a frequency very close to the resonance frequency of the DBE cavity. 
In summary, the onset of a DBE oscillator is the existence of a cavity with a DBE and the supersynchronization mechanism. Therefore, the SWS needs to support at least four modes (propagating and evanescent). The SWS can also be overmoded, as long as there are four modes that create the DBE, as it has been investigated in \cite{Maclachlan2024sub,MacLachlan2022Efficient}.

\section{DBE Oscillator Design}
\subsection{Dispersion Relation and Geometry}

The performance of the DBE oscillator is fundamentally governed by the unique dispersion characteristics of its cold SWS, illustrated in Fig. \ref{fig:disp_diagr_DBEO}. This dispersion diagram, obtained via eigenmode analysis using CST Studio Suite's eigenmode solver with the walls of the SWS unit cell modeled using perfect electric conductor (PEC) material, reveals the presence of a DBE at $\beta_{\mathrm{c}} d/\pi = 2$, corresponding to a DBE frequency of $f_\mathrm{d} = 50.143$ GHz. 
Notably, the use of glide symmetry (GS) in the waveguide design suppresses space harmonic coupling at $\beta_{\mathrm{c}} d/\pi = 1,3,...$, effectively preventing the formation of bandgaps at these intersection points due to the distinct parity of field components across harmonics \cite{Zvonimir_2024}. This property of GS ensures a smooth modal transition, which is critical for robust beam–wave interaction.

A key feature that enhances the oscillator’s performance is the selective interaction between the electron beam and the degenerate four waves in the proximity of the DBE point. The beamline, shown as a red dashed line in the figure and corresponding to a beam velocity of $u_0 = 0.344c$ (equivalent to an accelerating voltage of $33.22$ kV), intersects the branch with the DBE (solid blue for $\beta_{\rm{c}} d/\pi < 2$) just below the DBE point at $\beta_{\mathrm{c}} d/\pi = 2$. Importantly, field analysis at the beam position indicates that the mode relative to the other branch (dashed gray branch for $\beta_{\rm{c}} d/\pi < 2$) lacks an axial electric field component $E_z$, which is necessary for effective beam interaction. In contrast, the mode described by the branch with the DBE supports a nonzero $E_z$, thereby enabling energy exchange with the electron beam while suppressing coupling with the backward mode \cite{Castro23}. 



\begin{figure}[htpb]
\centering
\includegraphics[width=0.94\linewidth]{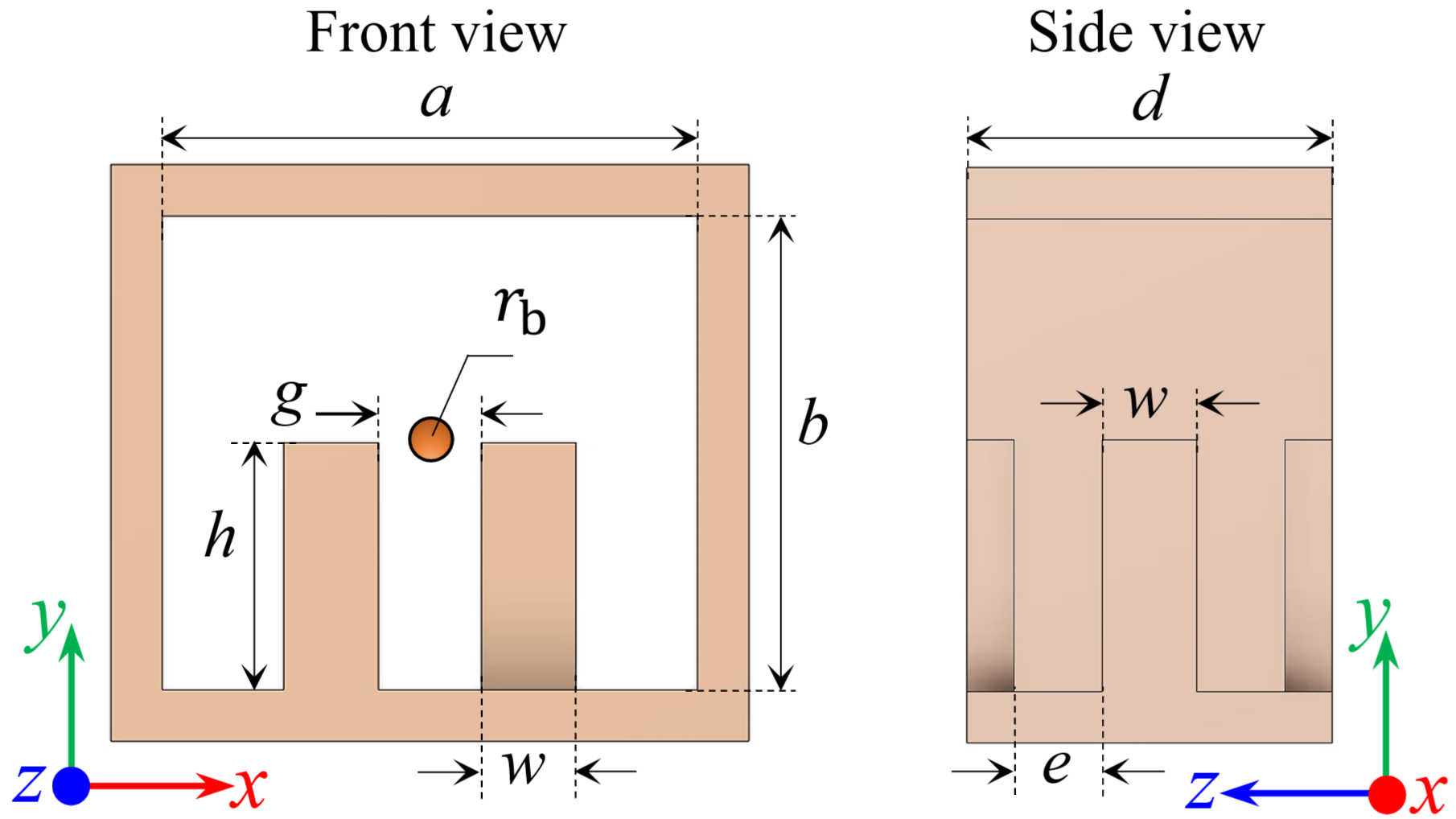}
\caption{Metallic unit cell of the multimodal SWS with glide symmetric corrugations. The cross-section of the electron beam of radius $r_{\text{b}}$ is shown in the front view. The ``cold" SWS supports a DBE.} 
\label{fig:DBE_UC_geometry}
\end{figure}

The unit cell geometry of the DBE oscillator structure is illustrated in Fig. \ref{fig:DBE_UC_geometry}. This design consists of a rectangular waveguide with two pillars of square cross section, with a shift of $d/2$ between them along the $z$-axis to ensure GS, and is based on previous designs reported in \cite{MineoDouble2010, PaoloniTHzBackward2016,Basu2021, Castro23}. The electron beam cross-section is shown in the front view to demonstrate its position relative to the metallic structures, ensuring alignment with the maximum electric field regions for effective interaction. The symmetry and spatial configuration of the unit cell are crucial for inducing a degenerate band-edge condition, which enhances field confinement and enables low-threshold oscillation. Additionally, the precise placement of the pillars within the waveguide is key to enforcing GS and achieving the backward mode coupling suppression discussed earlier. The dimensions are listed in Table \ref{tab:DBEO_UC dims}.

\begin{table}[H]
\caption{Dimensions of the DBE oscillator unit cell in the interaction section.}
\centering
\begin{tabular}{|>{\centering\arraybackslash}p{1.25cm}|>{\centering\arraybackslash}p{0.45cm}|>{\centering\arraybackslash}p{0.45cm}|>{\centering\arraybackslash}p{0.45cm}|>{\centering\arraybackslash}p{0.45cm}|>{\centering\arraybackslash}p{0.45cm}|>{\centering\arraybackslash}p{0.45cm}|>{\centering\arraybackslash}p{0.45cm}|>{\centering\arraybackslash}p{0.45cm}|>
{\centering\arraybackslash}p{0.45cm}|>
{\centering\arraybackslash}p{0.45cm}|}
\hline
\textbf{Dimension} & \textbf{$a$} & \textbf{$b$} & \textbf{$d$} & \textbf{$e$} & \textbf{$g$} & \textbf{$h$} & \textbf{$w$} & \textbf{$r_\mathrm{b}$} \\ 
\hline
\textbf{Val. (mm)} & 2.6 & 2.3 & 1.97 & 0.525 & 0.5 & 1.2 & 0.46 & 0.1 \\ 
\hline
\end{tabular}
\label{tab:DBEO_UC dims}
\end{table}

\begin{figure}[htpb]
\centering
\includegraphics[width=0.94\linewidth]{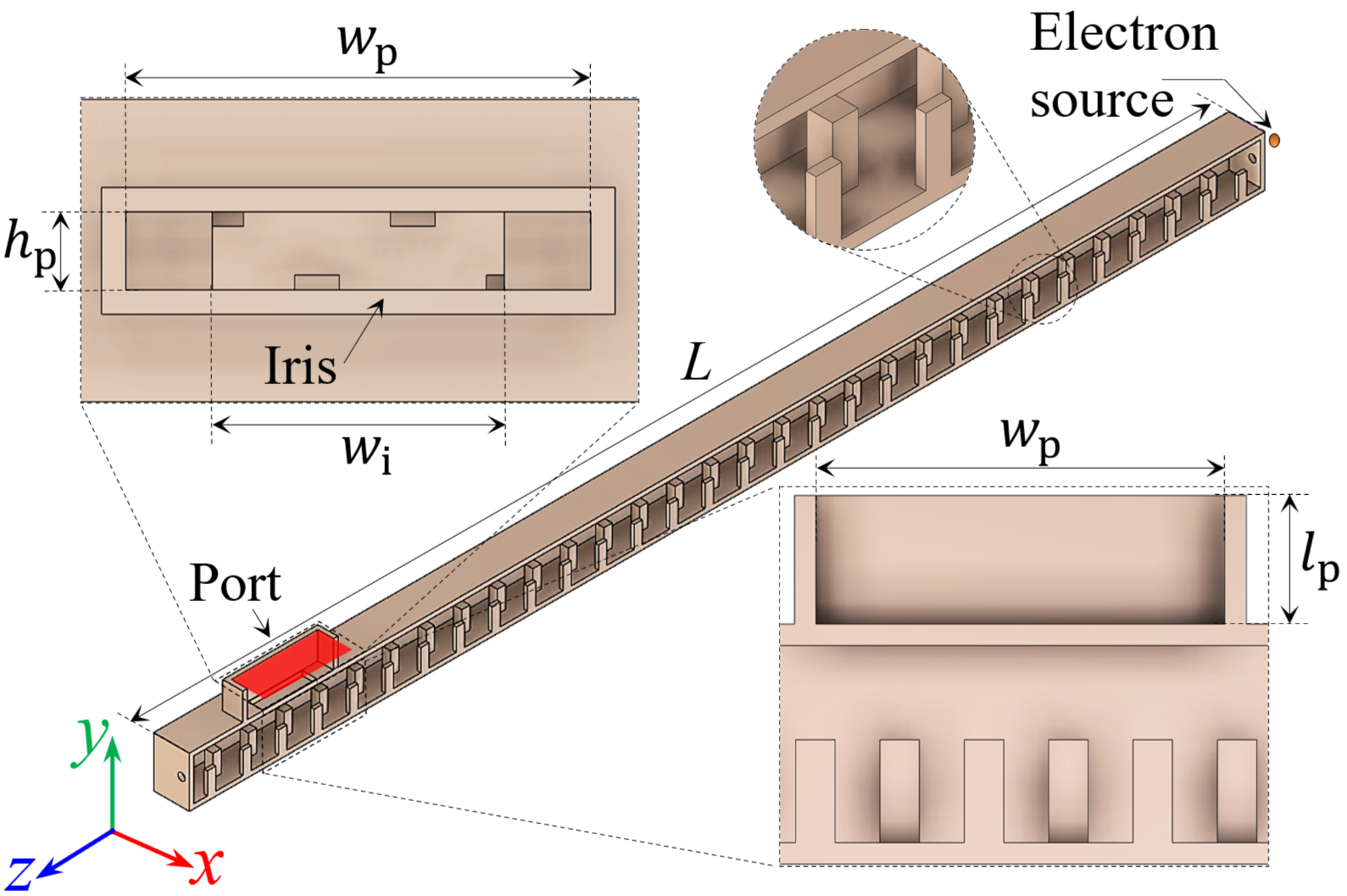}
\caption{Longitudinal cross section of the SWS for the DBE oscillator, which consists of the interaction zone and the coupler for power extraction at the rectangular port (red). The interaction zone is glide symmetric. The dimensions of the port and the iris used as a coupler are shown.} 
\label{fig:DBEO_full_str}
\end{figure}

Fig. \ref{fig:DBEO_full_str} presents the full longitudinal view of the SWS used in the DBE oscillator, including the interaction section, with $L=Nd$ unit cells, and the output coupler. The electron source is placed at the right end of the structure (top right corner), emphasizing the electron beam's injection into the interaction region. The meticulous arrangement ensures that the beam interacts coherently with the electromagnetic mode over multiple unit cells, enabling the build-up of oscillations close to the DBE frequency. At the left end of the structure, power is extracted through a port connected to a standard WR-19 rectangular waveguide. To enable efficient coupling between the oscillator and the external waveguide, an iris coupler is employed, as highlighted in the inset. This iris consists of a symmetric metal diaphragm inserted into the waveguide, introducing a discontinuity that facilitates the transfer of electromagnetic energy. Oriented so that its vertical edges are parallel to the electric field, aligned along the $x$-direction, the iris selectively excites transverse electric (TE) modes. The dimensions of the port and iris, and the length for $N=30$ unit cells, are listed in Table \ref{tab:DBEO_FullStr_dims}. 


\begin{table}[H]
\caption{Dimensions of the DBE oscillator coupler.}
\centering
\begin{tabular}{|>{\centering\arraybackslash}p{1.3cm}|>{\centering\arraybackslash}p{0.6cm}|>{\centering\arraybackslash}p{0.6cm}|>{\centering\arraybackslash}p{0.6cm}|>{\centering\arraybackslash}p{0.6cm}|>{\centering\arraybackslash}p{0.6cm}|>{\centering\arraybackslash}p{0.6cm}|>{\centering\arraybackslash}p{0.6cm}|>{\centering\arraybackslash}p{0.6cm}|>{\centering\arraybackslash}p{0.6cm}|}
\hline
\textbf{Dimension} & \textbf{$w_\mathrm{p}$} & \textbf{$h_\mathrm{p}$} & 
\textbf{$l_\mathrm{p}$} & \textbf{$w_\mathrm{i}$}& \textbf{$L$}\\ 
\hline
\textbf{Val. (mm)} & 4.7752 & 0.8 & 1.5 & 3 & 59.1\\ 
\hline
\end{tabular}
\label{tab:DBEO_FullStr_dims}
\end{table}

\subsection{Interaction Impedance}

The interaction impedance characterizes the strength of coupling between a guided RF wave and an electron beam by evaluating the longitudinal electric field component, $E_z$, at the beam's position \cite{pierce-twt50}. It is formally expressed as \cite{gewartowski65CH10}

\begin{equation}
    Z_{\mathrm{P}} = \frac{|E_{z}|^2}{2\beta_\mathrm{c}^2P}     \label{eqn:interaction_impedance}
\end{equation}

\noindent where $\beta_{\rm{c}}$ denotes the phase constant of the mode in the cold SWS, and $P$ is the time-averaged power carried by the mode \cite{pierce-twt50}. Here, $E_{z}$ represents the phasor of the longitudinal electric field component associated with the first spatial harmonic of the guided mode, i.e., in our case, it is the one associated with $\beta_{\rm{c}}$  near $2\pi$.

\begin{figure}[htpb]
\centering
\includegraphics[width=0.98\linewidth]{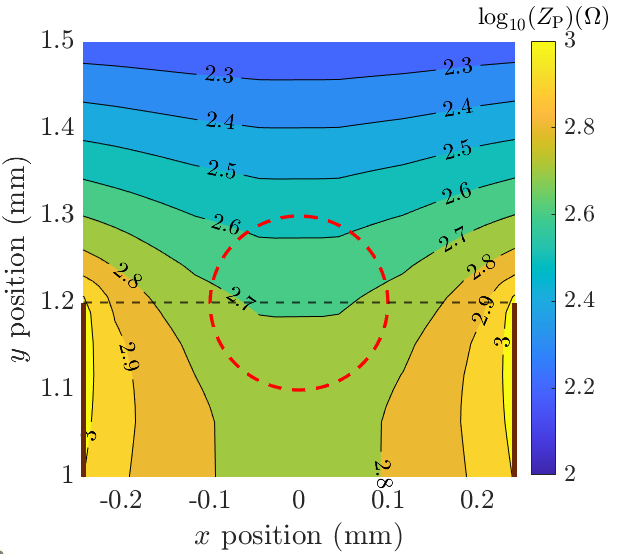}
\caption{Interaction impedance over the cross-section of the electron beam (inside the dashed-red circle) at $f=49.99$ GHz and $\beta_{\rm{c}} d/\pi = 1.6$. } 
\label{fig:DBE_Zn_surface}
\end{figure}

The spatial distribution of the interaction impedance $Z_{\mathrm{P}}$ within the region between the two inner pillars of the cold structure at an operating frequency of $49.98$ GHz is shown in Fig. \ref{fig:DBE_Zn_surface}. The frequency is chosen to be the one at which the system exhibits steady-state oscillations, as it will be clear later on in the paper. The color scale indicates the base-10 logarithm of $Z_{\mathrm{P}}$, highlighting the zones where beam–wave interaction is most effective. The highest impedance values, shown in yellow, are concentrated near the two pillars that extend symmetrically on both sides up to 1.2 mm along the $y$-axis (black vertical lines), where the electric field intensity is maximized. The red dashed circle denotes the cross-section of the electron beam, where the interaction impedance ranges between $400$ $\Omega$ and $630$ $\Omega$. This high impedance within the beam region plays a critical role in facilitating efficient energy exchange between the guided mode and the synchronized electron beam.


\section{Performance of the DBE Oscillator}

This section provides a detailed assessment of the output characteristics of the DBE oscillator, based on both time- and frequency-domain analyses. Full-wave particle-in-cell (PIC) simulations are employed to examine the oscillator’s behavior under optimal operating conditions identified from the cold-structure dispersion analysis. The goal is to demonstrate the onset of oscillation, quantify the resulting output power, and evaluate the spectral purity of the generated signal. 



\begin{figure}[htpb]
\centering
\includegraphics[width=1\linewidth]{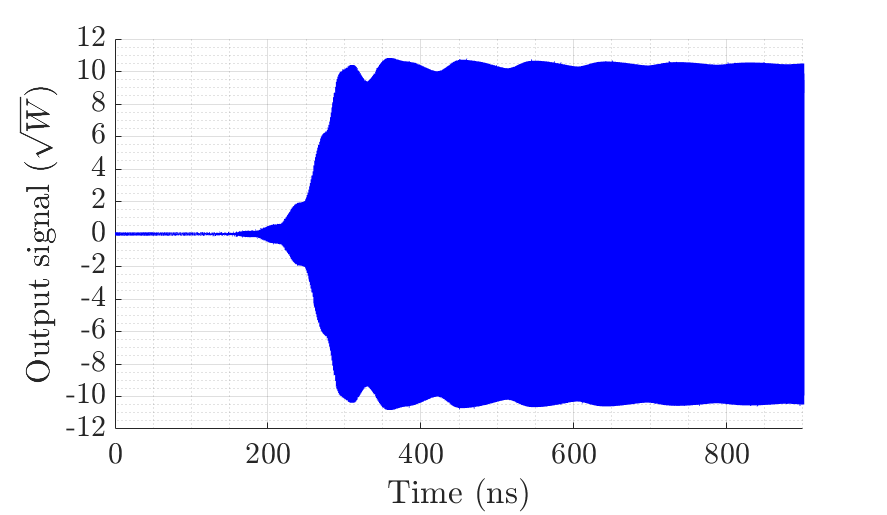}
\caption{Time domain signal at the output of the DBE oscillator with $N=30$ unit cells, $I_0 = 37$ mA, and $V_0=33.2$ kV. The steady oscillation
regime is reached at approximately $400$ ns, and the output power $P_\mathrm{o}$ is 55 W.} 
\label{fig:DBEO_Output_TD}
\end{figure}

Fig. \ref{fig:DBEO_Output_TD} shows the time-domain output signal obtained from a PIC simulation of the DBE oscillator using the structure composed of 30 unit cells modeled with pure copper. The simulation conditions include an injected electron beam current of $I_0 = 37$ mA and a beam voltage of $V_0 = 33.2$ kV, corresponding to the optimal beamline shown in Fig. \ref{fig:disp_diagr_DBEO}. The beam is confined by a magnetic field of around $B_z = 0.59$ T, which is five times the Brillouin magnetic flux density for this beam configuration \cite{Gilmour11}. The output waveform demonstrates the typical behavior of a startup transient followed by a steady-state oscillation regime. After an initial build-up period of several hundred nanoseconds, a nearly sinusoidal signal with a stable amplitude emerges, indicating the onset of self-sustained oscillation. The output power is calculated to be approximately $55$ W, demonstrating the high efficiency of the beam–wave interaction under DBE conditions.

\begin{figure}[htpb]
\centering
\includegraphics[width=0.98\linewidth]{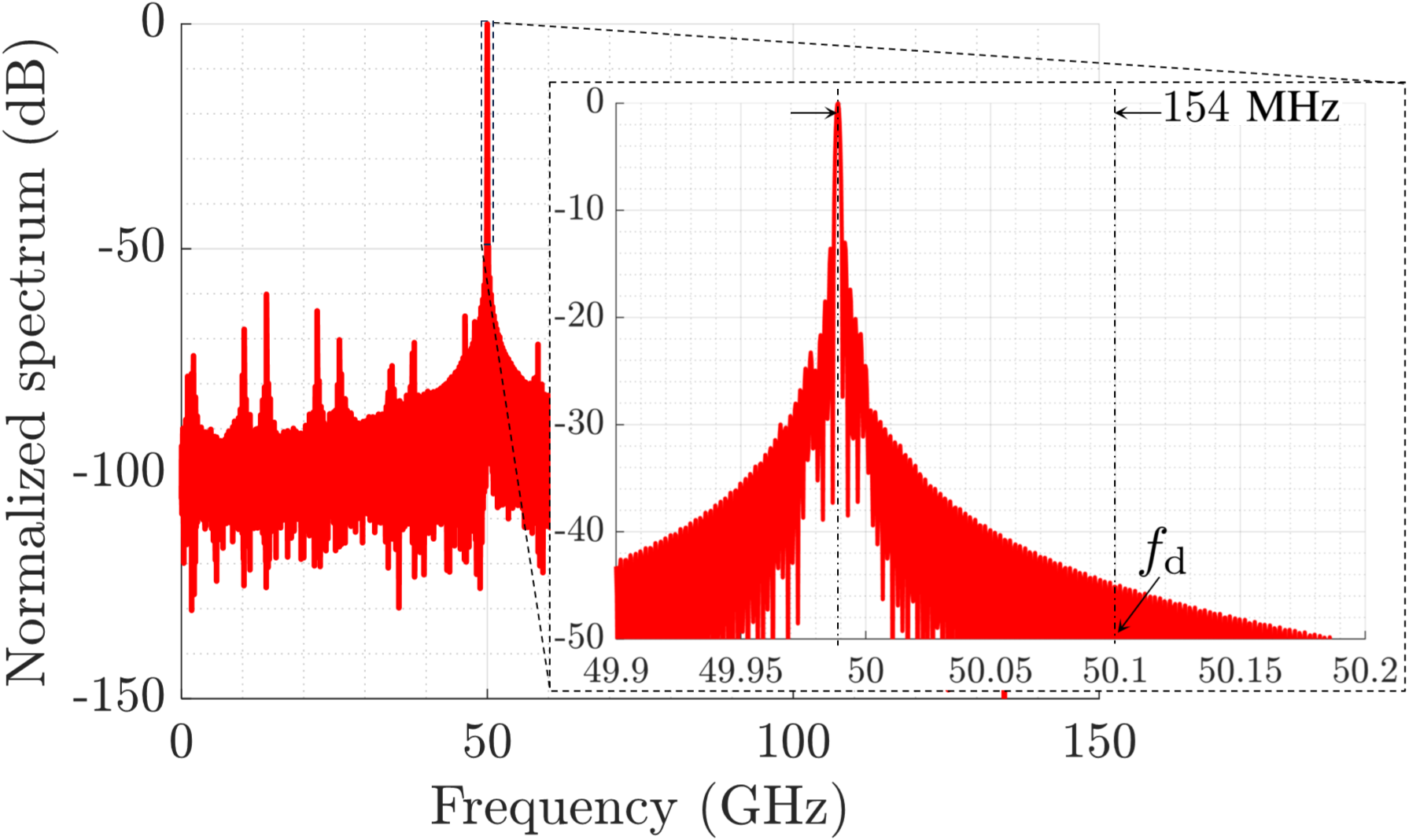}
\caption{Normalized spectrum of the signal shown in Fig. \ref{fig:DBEO_Output_TD} of the DBE oscillator. The time window for calculating the spectrum is $400$ to $900$ ns. The oscillation frequency of $49.989$ GHz is $154$ MHz lower than the DBE frequency, as expected. }
\label{fig:DBEO_Output_FD}
\end{figure}

Fig. \ref{fig:DBEO_Output_FD} displays the normalized frequency-domain spectrum of the time signal shown in Fig. \ref{fig:DBEO_Output_TD}, revealing detailed spectral features of the oscillator's output. The main spectral peak appears slightly below the cold DBE frequency, $f_\mathrm{d}=50.143$ GHz, with an observed center frequency of approximately $49.989$ GHz. The inset further emphasizes the spectral purity and narrow width and highlights a frequency separation of $154$ MHz with respect to $f_\mathrm{d}$, indicative of a stable, single-mode operation. This shift can be attributed to beam loading effects, which modify the phase velocity of the mode due to space-charge interactions and energy exchange. The narrow spectral width 
confirms the presence of a highly coherent oscillation mode supported by the structure. 

\section{Parametric Analysis of the DBE Oscillator}

To evaluate the performance and stability of the DBE oscillator under varying operating conditions, this section presents a comprehensive parametric analysis. We examine the effects of the beam current $I_0$, electron beam velocity $u_0$, and structure length on key performance metrics, including the output power $P_\mathrm{o}$, electronic efficiency $\eta$, and the stability of the oscillation frequency $f$. These parameters are crucial to assess the practical feasibility of the DBE oscillator and its robustness in real-world scenarios. 

\begin{figure}[htpb]
\centering
\includegraphics[width=0.98\linewidth]{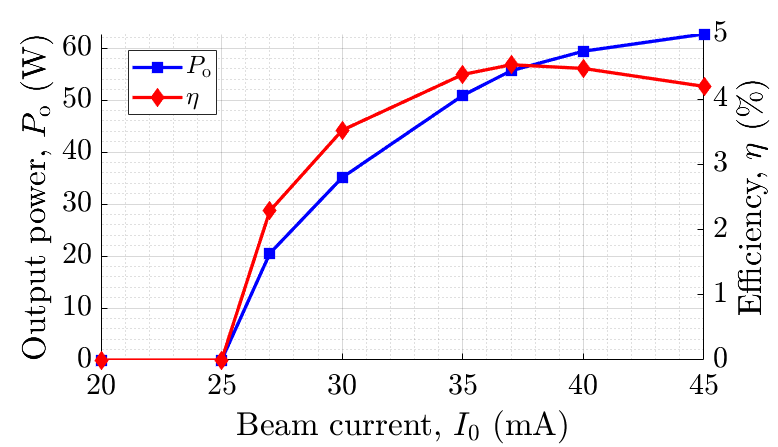}
\caption{Output power and efficiency as a function of the beam current, with $u_0=0.344\mathrm{c}$. The efficiency is maximum for a beam current of around $37$ mA, and the power saturates at around $63$ W.} 
\label{fig:DBEO_Po_Eff_vs_I0}
\end{figure}

Fig. \ref{fig:DBEO_Po_Eff_vs_I0} presents the output power $P_o$ and the electronic efficiency $\eta$ of the DBE oscillator as functions of the beam current $I_0$, for an electron beam velocity corresponding to $u_0 = 0.344c$ and a structure with 30 unit cells. The plot shows that the beam current threshold lies slightly above $25$ mA, at which point the output power remains negligible. Beyond this threshold,  as the beam current increases, the output power grows rapidly after a beam current of about $27$ mA, and then saturates after $63$ W (which corresponds to a beam current greater than $45$ mA).
On the other hand, the efficiency, which is defined as $\eta = P_o/(I_0V_0)$, reaches a peak of about $4.5 \%$ at $I_0 = 37$ mA, after which it slightly decreases, likely due to saturation effects and nonlinear beam–wave interaction dynamics. 

\begin{figure}[htpb]
\centering
\includegraphics[width=1\linewidth]{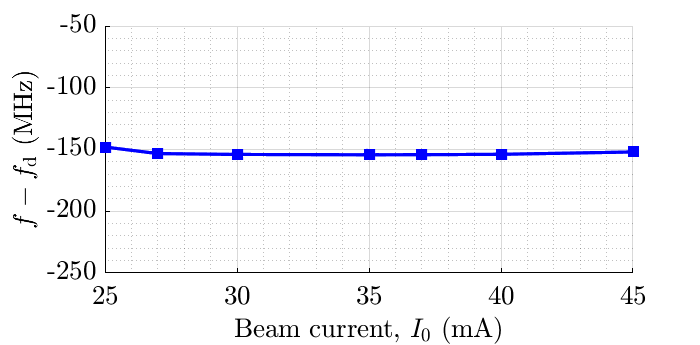}
\caption{The oscillation frequency, $f$
remains approximately constant as the beam current $I_0$ varies. A difference of approximately 154 MHz between the oscillation frequency and the DBE frequency $f_{\mathrm{d}}$ is maintained over a large range of current variation. We assume $u_0=0.344\mathrm{c}$.} 
\label{fig:DBEO_FreqShift_I0}
\end{figure}

Fig. \ref{fig:DBEO_FreqShift_I0} provides the oscillation frequency $f$ relative to the DBE frequency $f_\mathrm{d}$ as a function of the beam current $I_0$. The plot shows that $f$ remains approximately constant around 154 MHz below $f_\mathrm{d}$, across the entire range of beam currents examined, from $27$ mA to $40$ mA. At $25$ mA, although the simulated frequency is closer to the DBE one, the oscillator has not yet reached the threshold current, and the output power is negligible, as shown earlier. This behavior indicates that the oscillation frequency is largely insensitive to changes in the beam current within this range above threshold, which is a desirable characteristic for stable operation. The relative flatness of the oscillation frequency plot suggests that once the oscillator reaches the threshold for sustained oscillations, further increases in beam current primarily affect the output power rather than the frequency. This stability in $f$ is consistent with the expected properties of the DBE oscillation mechanism, where the interaction dynamics are dominated by the flat dispersion near the degeneracy point.

\begin{figure}[htpb]
\centering
\includegraphics[width=0.98\linewidth]{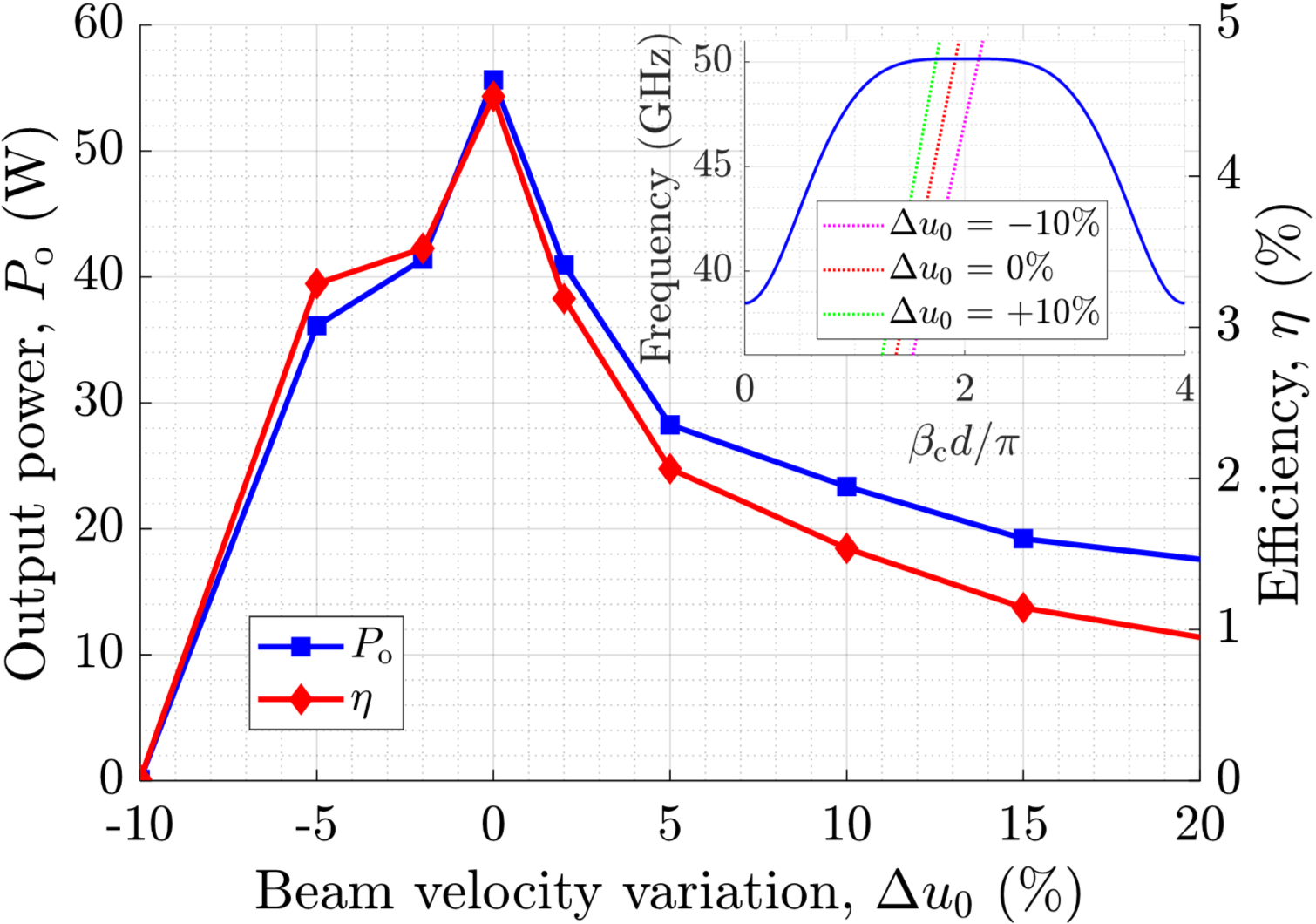}
\caption{Output power and efficiency as a function of the electrons' velocity. We assume $I_0=37$ mA. Power and efficiency are kept at high levels for a large variation of the electrons' velocity $u_0$ because of the flatness dispersion diagram at the DBE. The efficiency is maximum at $u_0=0.344\mathrm{c}$.}

\label{fig:DBEO_Po_Eff_vs_u0}
\end{figure}

The performance of the DBE oscillator as a function of the electron beam velocity is shown in Fig. \ref{fig:DBEO_Po_Eff_vs_u0}, where both the output power $P_o$ and the electronic efficiency $\eta$ are plotted against the beam velocity variation $\Delta u_0$, expressed as a percentage deviation from the nominal velocity corresponding to the DBE condition ($u_0 = 0.344c$). The results reveal that the oscillator is highly sensitive to the beam velocity: the output power and efficiency sharply increase as $\Delta u_0$ approaches zero, reaching their maximum values at $\Delta u_0 = 0\%$. This confirms that maximum energy transfer from the electron beam to the RF field occurs precisely at the DBE synchronization point. For deviations above or below this optimal velocity, the performance degrades, with the output power and efficiency progressively decreasing, underscoring the importance of precise velocity tuning in DBE-based oscillators.

\begin{figure}[htpb]
\centering
\includegraphics[width=1\linewidth]{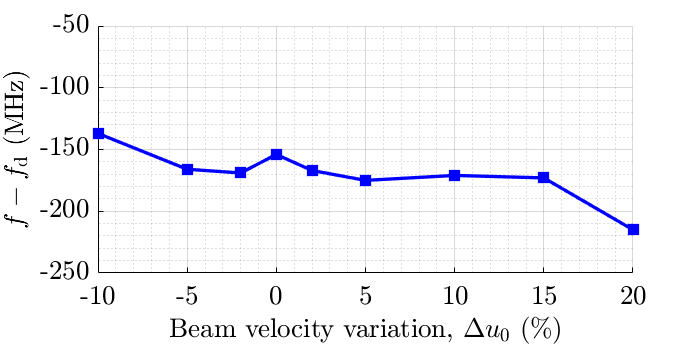}
\caption{The frequency of oscillation of the DBE oscillator, $f$, remains approximately constant when varying the electron's velocity $u_0$ over a wide range. This property is provided by the flatness of the dispersion diagram at the DBE. We assume $I_0=37$ mA.} 
\label{fig:DBEO_FreqShift_u0}
\end{figure}

Additionally, Fig. \ref{fig:DBEO_FreqShift_u0} presents the oscillation frequency relative to the DBE frequency, $f-f_\mathrm{d}$, as a function of the beam velocity variation $\Delta u_0$. The plot indicates that the frequency remains more or less constant and relatively close to the DBE frequency, demonstrating the oscillator’s frequency stability near the supersynchronism point when varying the beam velocity. Notably, at $\Delta u_0 = 0\%$, the oscillation frequency is about $154$ MHz below $f_\mathrm{d}$, a point that also corresponds to the maximum output power $P_o$ and electronic efficiency $\eta$, as seen in Fig. \ref{fig:DBEO_Po_Eff_vs_u0}. However, for larger deviations in velocity, the frequency difference becomes slightly more pronounced, especially beyond $\Delta u_0 = 15\%$, where $f-f_\mathrm{d}$ exceeds $-200$ MHz. The slight decrease in oscillation frequency when the beam velocity increases is explained by observing that the synchronization point moves further on the left in the inset of Fig.~\ref{fig:DBEO_FreqShift_u0}, hence it occurs at a slightly lower frequency than the DBE. 

\begin{figure}[htpb]
\centering
\includegraphics[width=0.98\linewidth]{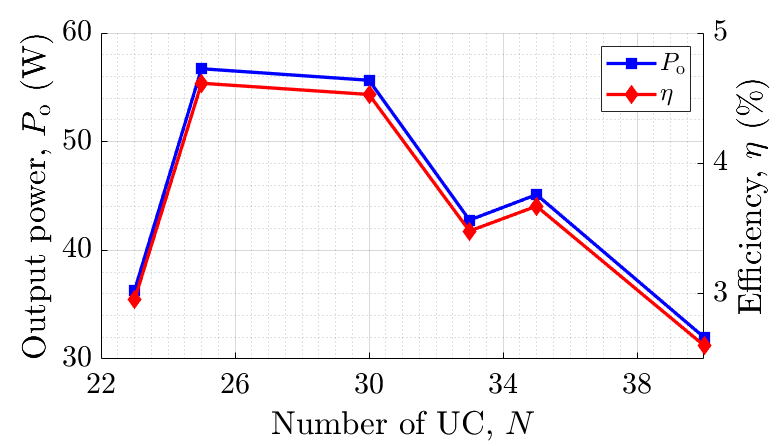}
\caption{Output power and efficiency as a function of the SWS length $N$, assuming $I_0=37$ mA and $u_0=0.344c$. 
}
\label{fig:DBEO_Po_Eff_vs_N}
\end{figure}

Fig. \ref{fig:DBEO_Po_Eff_vs_N} illustrates the output power $P_o$ and efficiency $\eta$ of the DBE oscillator as functions of structural length variation, expressed through the change in the number of unit cells $N$. In all cases, the beam parameters are fixed at $I_0 = 37\,\text{mA}$ and $u_0 = 0.344c$. As observed, both output power and efficiency remain nearly constant when a few unit cells are removed ($N < 30$), indicating some tolerance to minor reductions in structure length. However, increasing the number of unit cells ($N > 30$) results in a noticeable degradation in performance, with $P_o$ decreasing from approximately $58$ W to $32$ W and efficiency dropping from around $4.5\%$ to $2.5\%$. 


\begin{figure}[htpb]
\centering
\includegraphics[width=1\linewidth]{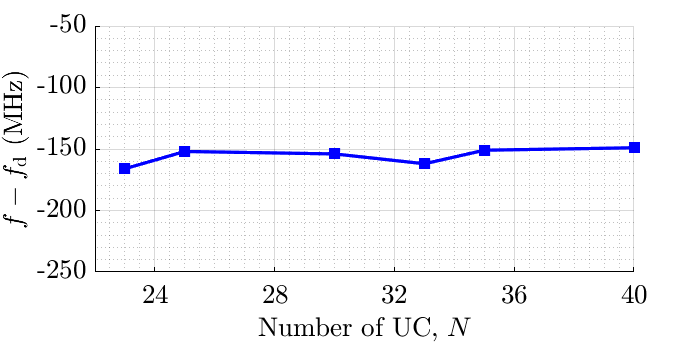}
\caption{The frequency of oscillation of the DBE oscillator, $f$, remains approximately constant at 49.98 GHz when varying the SWS length. We assume $I_0=37$ mA and $u_0=0.344c$.}
\label{fig:DBEO_FreqShift_N}
\end{figure}

Fig. \ref{fig:DBEO_FreqShift_N} shows the frequency difference $f-f_\mathrm{d}$ between the actual oscillation frequency and the DBE frequency, as a function of $N$. The results indicate that the frequency remains relatively stable, with an average deviation of approximately $154$ MHz and a maximum variation of up to $15$ MHz across the different cases.

\section{Conclusion}
We presented the design and performance analysis of a degenerate band-edge oscillator based on a double-corrugated slow-wave structure that supports a four-mode supersynchronization. The unique dispersion properties of the DBE, characterized by vanishing group velocity and high density of electromagnetic modes, were harnessed to achieve self-sustained oscillations with relatively low starting beam current and high interaction impedance. Through full-wave PIC simulations, we demonstrated stable single-frequency oscillations near the DBE resonance for a 30-unit-cell structure, delivering an output power of approximately $55$ W at $49.989$ GHz. The results also highlighted the role of glide symmetry in ensuring modal isolation and maximizing beam–wave energy exchange by eliminating coupling with a backward wave.

Importantly, we performed a comprehensive parametric study to evaluate the robustness and scalability of the DBE oscillator. Our analysis confirmed that the oscillation frequency remains highly stable against variations in beam current, electron beam velocity, and structural length, a direct consequence of the extremely flat dispersion at the DBE. While performance was sensitive to deviations from the optimal synchronism point, particularly in beam velocity, the oscillator maintained efficient operation within a practical range of parameter fluctuations. These findings validate the DBE oscillator as a promising candidate for high-efficiency, coherent radiation sources at millimeter-wave frequencies and pave the way for further development of DBE-based devices using alternative geometries, advanced fabrication techniques, and integration into compact vacuum electronic systems.

\section*{Acknowledgment}


The authors are thankful to DS SIMULIA for providing CST Studio Suite, which was instrumental in this study. 

\appendix 
\section*{DBE Multimode Bloch analysis} \label{Appendix:A}

We report the dispersion diagram of the lossless waveguide whose unit cell is that of Fig. \ref{fig:DBE_UC_geometry}, considering also the evanescent modes that degenerate at the DBE frequency observed in Fig. \ref{fig:disp_diagr_DBEO}. The complex dispersion diagram is computed using a multi-mode (MM) Bloch analysis, as shown in \cite{Mesa_2021_MAP_Simulation_assisted_dispersion, Giusti_2022_TAP_Multimode}. Using full-wave simulations with waveguide ports at the boundaries ($z = 0$ and $z = d$)  of the unit cell  shown in Fig. \ref{fig:DBE_UC_geometry}, the transfer $\mathbf{T}$ matrix is obtained. Applying the Floquet theorem leads to   the following eigenvalue problem: 

\begin{equation}
    \mathbf{T} 
    \begin{bmatrix} 
        \mathbf{V}_2 \\ \mathbf{I}_2 
    \end{bmatrix}
    =
    e^{- j k d} 
    \begin{bmatrix} 
        \mathbf{V}_1 \\ \mathbf{I}_1 
    \end{bmatrix} \label{eqn:MM-Bloch}
\end{equation}
where $\mathbf{V}$ and $\mathbf{I}$ at each port are defined as

\begin{equation}
    \mathbf{V} =
    \begin{bmatrix}
        V^{(1)} \\
        V^{(2)} \\
        \vdots \\
        V^{(M)}
    \end{bmatrix},
    \quad
    \mathbf{I} =
    \begin{bmatrix}
        I^{(1)} \\
        I^{(2)} \\
        \vdots \\
        I^{(M)}
    \end{bmatrix}. 
\end{equation}

Here, \( m = 1, 2, \dots, M \) indexes the port modes defined at each end of the unit cell between \( z = 0 \) and \( z = d \), and \( k = \beta_c - j\alpha_c \) is the complex wavenumber associated with the set of \( M \) eigenmodes. The use of the transfer matrix $\mathbf{T}$ allows for the accurate determination of the Bloch mode wavenumbers, based on the fact that the port modes form an orthonormal basis used to represent the Bloch modes.

\begin{figure}[htpb]
\centering
\includegraphics[width=1\linewidth]{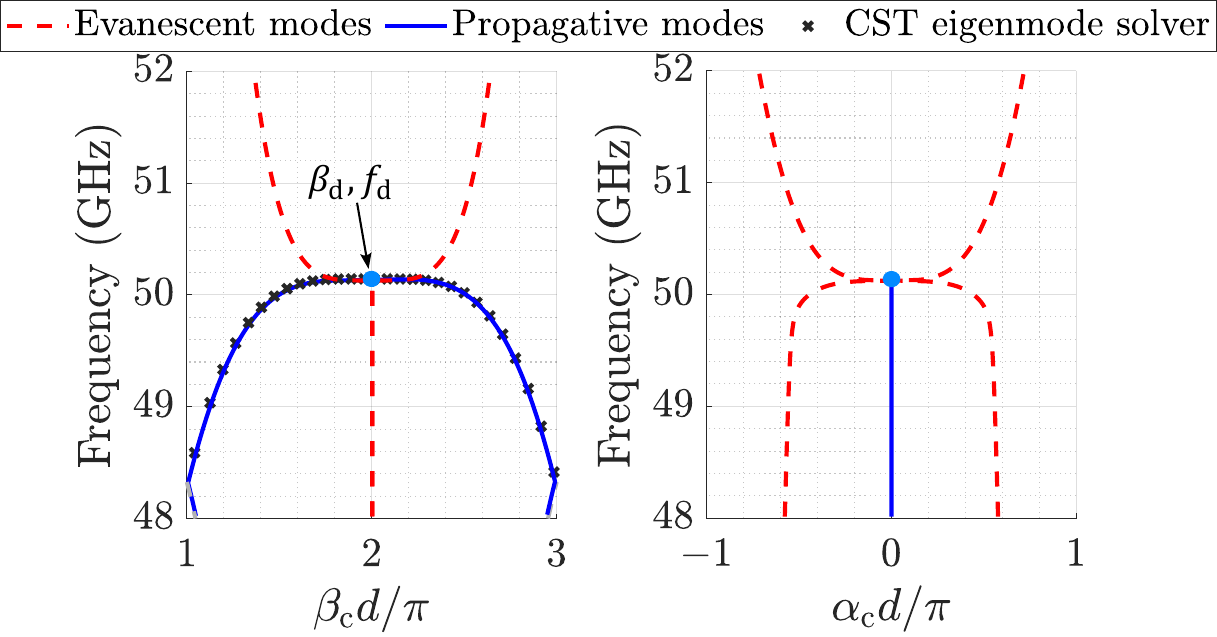}
\caption{Cold dispersion diagram of the modes in the SWS with DBE. We plot the fours modal branches above and below $f_\mathrm{d}$, including those with complex wavenumber $k=\beta_\mathrm{c}-j\alpha_\mathrm{c}$. The dashed red curves are evanescent modes that degenerate at the DBE point ($\beta_{\textrm{d}},f_\textrm{d}$), the solid blue line are propagative modes, also coalescing at  ($\beta_{\textrm{d}},f_\textrm{d}$); they are found also using the CST Eigenmode Solver (black crosses). }
\label{fig:disp_diagram_MM}
\end{figure}

In Fig. \ref{fig:disp_diagram_MM} the complex wavenumbers are shown. There is an excellent agreement between the real-wavenumber solution obtained using (\ref{eqn:MM-Bloch}) and that from the CST Eigenmode Solver. This analysis allows us to observe how the coalescence of four evanescent and propagative modes leads to the flat band observed for the DBE.

\bibliographystyle{IEEEtran}
\bibliography{Bibliography_DBE_Oscillator}

\end{document}